\documentclass[aps,preprintnumbers,amsmath,
amssymb,a4,floatfix,preprint]{revtex4}

\usepackage{graphicx}
\usepackage{epsfig}
\usepackage{bm}

\begin{document}
\bibliographystyle{prsty} 
\title{Exact transformation for spin-charge separation of spin-$\frac{1}{2}$ Fermions
without constraints
}
\author{Stellan \"Ostlund}  
\email{ostlund@fy.chalmers.se}
\author{Mats Granath}
\email{mgranath@fy.chalmers.se}
\affiliation{
G\"oteborgs Universitet\\ Gothenburg 41296, Sweden \\ 
}

\def\ssection#1{\section*{#1}}
\newcommand{\jay}{J}
\newcommand{\cnodot}{}
\newcommand{\nb}[1]{n^{b}}
\newcommand{\nf}[1]{n^{f}}
\newcommand{\cdop}[2]{c^{\dagger}_{#2} }           
\newcommand{\cop}[2]{c^{\phdagger}_{#2 }}
\newcommand{\bdop}[1]{b^{\dagger}_{} }           
\newcommand{\bop}[1]{b^{\phdagger}_{}}
\newcommand{\chdop}[1]{c^{\dagger}_{} }           
\newcommand{\chop}[1]{c^{\phdagger}_{}}
\newcommand{\Q}[1]{Q^{\phdagger}_{#1}}
\newcommand{\Qd}[1]{Q^{\dagger}_{#1}}
\newcommand{\phdagger}{\phantom{\dagger}}
\newcommand{\subs}[2]{ {#1}_{#2} }
\newcommand{\imag}{i}
\newcommand{\sups}[2]{ {#1}^{#2} }
\newcommand{\expectation}[1]{\langle \;  #1 \; \rangle}
\newcommand{\nh}{\hat{n}}
\newcommand{\rp}{{r'}}
\newcommand{\hopi}[1]{\hat{h}_{#1} }
\newcommand{\hop}{{h}}
\newcommand{\half}{{\textstyle \frac{1}{2}}}
\newcommand{\e}[1]{{ \langle\,#1\,\rangle}}
\newcommand{\ket}[1]{{ \, | \,  { #1 }  \, \rangle}}
\newcommand{\bra}[1]{{\langle\, {#1} \, | }}
\newcommand{\braket}[2]{{\langle\,#1\,|\,#2 \, \rangle\,}}
\newcommand{\quarter}{{\textstyle \frac{1}{4}}}
\newcommand{\dmu}{{\delta_\mu}}
\newcommand{\down}{\downarrow}
\newcommand{\up}{\uparrow}
\newcommand{\Hhub}{ H}
\newcommand{\bS}{\bm{S}}
\newcommand{\bP}{\bm{P}}
\newcommand{\bQ}{\bm{Q}}
\newcommand{\teezero}{{T_0}}
\newcommand{\teeplus}{{T_1}}
\newcommand{\teeminus}{{T_{-1}}}

\date{\today}

\begin{abstract}
We demonstrate 
an exact local transformation 
which maps a purely Fermionic manybody system  to a system of 
spinfull Bosons and
spinless Fermions, demonstrating a possible path to a non-Fermi liquid state.
We apply  this
to the half-filled Hubbard model and show how 
the 
transformation maps the ordinary spin half Fermionic degrees of freedom
exactly and without introducing Hilbert space constraints  to a charge-like 
``quasicharge'' fermion and a spin-like ``quasispin'' Boson while 
preserving all the 
symmetries of the model.  We present 
approximate solutions with localized 
charge which emerge naturally from the Hubbard model in this 
form. Our results 
strongly suggest that charge tends to remain localized
for large values of the Hubbard $ U $.

\end{abstract}

\maketitle
\renewcommand{\cdop}[2]{c^{\dagger}_{#1,\,#2}}           
\renewcommand{\cop}[2]{c^{\phdagger}_{#1,\, #2 }}
\renewcommand{\bdop}[1]{b^{\dagger}_{} }           
\renewcommand{\bop}[1]{b^{\phdagger}_{}}
\renewcommand{\chdop}[1]{{\hat{c}}^{\dagger}_{#1} }           
\renewcommand{\chop}[1]{{\hat{c}}^{\phdagger}_{#1}}
\renewcommand{\nb}[1]{n^{b}_{}}
\renewcommand{\nf}[1]{n^{f}_{}}
\newcommand{\gee}[2]{ q^{#1}_{#2 } }
\newcommand{\bgee}[1]{ {\bm{q} }_{\, #1}   }

The Hubbard model is a paradigm for one of the most difficult
quantum manybody systems; strong interactions competing
with the kinetic energy in a dense electron gas.
These problems 
must be explored either analytically by relatively poorly
controlled approximate methods or by 
methods such as Monte-Carlo, dynamical mean field theory (DMFT) or
finite size calculations.  A major difficulty is the absence of a framework
for such systems that competes in simplicity and rigor with
conventional Landau Fermi-liquid theory.

A particularly intriguing scenario for non-Fermi liquid behavior is 
``spin-charge separation'' with a low-energy effective theory where 
spin and charge excitations are decoupled. This idea 
has attracted widespread attention
because it migh provide  a route to
high-temperature superconductivity 
in a system with strong local repulsion. 
In 1-D it is known through the use of ``bosonization'' that spin-charge separatation does occur 
at low energy.\cite{LutherEmery} 
In higher dimensions,
much less is known.
The usual approach to spin charge separation is to construct 
electron operators as composite objects consisting of  
a charge part, the ``holon'', and a spin part, the ``spinon''. 
To avoid 
unphysical states 
contraints have to be included which are then
treated by approximate methods.\cite{fradkin}

Here we  shall construct
electron operators as composites of
charge-like and spin-like  operators
that do not give rise to any unphysical states 
and thus avoids introducing 
any constraints. An exact local mapping
which is valid in all dimensions will be shown to exist 
from a purely Fermionic manybody system to an equivalent system of 
Fermions and spin-like operators obeying Bosonic commutation relations
between different lattice sites. We shall  explore this  
with the 2-D Hubbard model, for which the resulting mapping bears a 
resemblance to the mappings of the local Hilbert space 
achieved by introducing spinons and holons.

\section*{Transformation to quasiparticle operators.}
Let $ n_r $ be the particle number at site $ r $, and
$ n_{\up} $ and $ n_{\down} $ be respectively the number
of up and down electrons: 
$ n_r =  n_{\up,r} +  n_{\down,r} $.
Since the Hubbard interaction obeys
$   n_{\downarrow, \, r } n_{\uparrow, \, r }  =  \half ( n_r - 1 
 +   (n_r - 1)^2  ) $, by choosing the chemical potential appropriately we can replace 
the usual Hubbard interaction with
the particle-hole symmetric interaction
$\half \,  U \, ( n_r - 1)^2 $.
We thus 
write the Hubbard model  in the 
particle-hole symmetric form $ 
\Hhub  = - t \sum_{\e{r\rp}} ( \cdop{s}{r} \cop{s}{\rp} \; + \;  \textstyle{ CC } ) + U \sum_r \half\, ( n_r - 1)^2
$
where the sum is over all sites and nearest neighbor bonds counted once.

We now demonstrate an exact transformation of the local
operator algebra of the Hubbard model to new Fermionic  $ CP^1 $
``quasicharge'' 
$ \chop{r} $ 
and  $ SU(2) $ ``quasispin'' operators 
$ \gee{i}{r} $ 
that obey, respectively, Fermi and Bose
statistics. The operators are given by
\begin{eqnarray} \label{eq:toquasi}
\chop{r} & = &  \cdop{\uparrow}{r}( 1 - n_{\downarrow,\,r}) + (-1)^r \cop{\uparrow}{r} n_{\downarrow , \, r} \\ 
\gee{+}{r} & = & ( \cdop{\uparrow}{r} - ( -1)^r \cop{\uparrow}{r} \; ) \; \cop{\downarrow}{r} \\ \nonumber
\gee{-}{r} & = &  (\gee{+}{r})^{\dagger} \\ \nonumber
\gee{z}{r}  & = &  \textstyle{\half}  -  n_{\downarrow,\,r} 
\end{eqnarray}
where $ -1^r $ is $ \pm 1 $ depending on which sublattice 
site $ r $ is on.  We define the $ x $ and $ y $ component of quasispin by 
$ \gee{x}{r} = \half \, ( \gee{+}{r} + \gee{-}{r} ) $ 
and $ \gee{y}{r} = \frac{1}{2 i } \, ( \gee{+}{r} - \gee{-}{r} ) $.
The factor appearing as $ (-1)^r $ can in fact be chosen as an arbitrary phase, but is
fixed here to preserve symmetries specific to the Hubbard model. 
This ''nonlinear'' transformation generalizes earlier work on
transformations that were strictly canonical, and hence did
not alter commutation relations of the fermi operators.\cite{mele,gunn}

The new operators obey the following algebra, which can be
verified  by straightforward 
manipulations:
$ \{ \chop{r}, \chdop{\rp} \} =  \delta_{r,\rp}$ ,
$ \{ \chdop{r}, \chdop{\rp} \} = 0 $ ,
$ \left[\, \chdop{r}, \gee{i}{\rp} \, \right] = 0 $ ,
$ \left[ \, \gee{i}{r}, \gee{j}{\rp} \right] = i \delta_{r\rp}  \sum_{k} \epsilon_{ijk} \gee{k}{r} $

We thus find that $ \gee{i}{r}  $, $ \chdop{r} $ and $ \chop{r}$ are 
respectively independent spin half bosonic and ``spinless'' Fermionic operators.
The usual operators are given by
\begin{eqnarray}
\cdop{\uparrow}{r} & =  & \chop{r} \; ( \, \half + \gee{z}{r} ) \;+ 
	\; (-1)^r \; \chdop{r} (\half - \gee{z}{r} ) \\ \nonumber
\cdop{\downarrow}{r}   & =  & \gee{-}{r} \, ( \,  \chop{r} \; - 
	\; (-1)^r \,  \chdop{r} \,  ).
\end{eqnarray}

We define the quasicharge operator as  
$ n^f_r  = \chdop{r} \chop{r} $. 
We find the following relations with the usual operators 
expressed in terms of ordinary Fermions:
$ n_r   =   1   - 2  n^f_r \gee{z}{r} $,
$ s^i_r  =   ( 1 - n^f_r \, ) \;  \gee{i}{r} $ and
$ n^f_r  =  (n_r - 1)^2  $
where $ s^i_r = \half \sum_{\alpha,\beta} \cdop{\alpha}{r} \sigma^i_{\alpha \beta } \cop{\beta}{r} $,
with $ \sigma^i  $ the Pauli matrices.
We also define the local ''pseudospin'' operators  
$ p^{i}_r = n^f_r \gee{i}{r} $ which are the generators of the SU(2) algebra which correponds to ''rotations'' between the empty and 
doubly occupied states.\cite{shiba72} The formal resemblance to spin
symmetry has led to the name pseudospin. The total z-component of pseudospin 
can therefore be seen to be half the number of doubly occupied
sites minus the number of empty sites, which is precisely the charge
relative to half filling. Pseudospin is a symmetry of the particle-hole symmetric 
Hubbard model but is broken by a chemical potential.

We note that, aside from phase factors, the quasicharge operators
always create even charged states from odd and vice versa, as they
must since they are Fermionic.
However, there is a precise phase relationship which is not so easily interpreted 
that is enforced by the commutation relations and symmetries.

Our Hubbard model can now be rewritten {\it exactly } in terms of the quasiparticle operators as
\begin{equation}
\label{eq:newhub}
H = t \;  ( \teezero + \teeplus + \teeminus ) + U h_U 
\end{equation} 
with $ h_U  =  \half  \sum_r  \, \chdop{r}\chop{r}  $ and
\begin{eqnarray} \label{eq:teepm}
\teezero & = & \frac{1}{2} \sum_{\e{r,\rp}}  \,  ( 1 + 4 \bgee{r}\cdot\bgee{\rp} )(\chdop{r}\chop{\rp} +CC) \\ \nonumber
\teeplus & = & \frac{1}{2} \sum_{\e{r,\rp}} -1^r \, ( 1 - 4 \bgee{r}\cdot\bgee{\rp} )(\chdop{r}\chdop{\rp} ) \\ \nonumber
\end{eqnarray}
and $T_{-1} = T_{1}^\dagger $.
In this representation, the symmetry under the entire $ SU(2) $ quasispin rotation is manifest  and we note 
that the Hubbard interaction becomes simply the quasicharge number operator.

We 
define total quasispin $ \bf{Q} $ , spin $ \bf{S} $  and 
pseudospin $ \bf P $  respectively
by $ \bQ  = \sum_{r} \bgee{r} $ etc. 
$ \bP,\bQ $ and $ \bS $ all obey the SU(2) algebra 
$ \left[ \, P_i , P_j  \right] = i \sum_{k} \epsilon_{ijk} P_k \; $
etc.
We see from  the definitions above Eq. \ref{eq:newhub} 
the relationship,
$ p^{i}_r = n^f_r q_r^i $ and 
$ s^i_r  =   ( 1 - n^f_r \, ) \;  \gee{i}{r} $. Thus
$ \bQ = \bP + \bS $, i.e. quasispin can be exactly split into
pseudospin and spin, with $ n^f_r $ and $ ( 1 - n^f_r) $ providing
the projection operator of $ \bQ $ into either spin or pseudospin.
Since $ \bQ $ commutes with $ \chdop{r} $ and the Hubbard Hamiltonian 
$ H $ 
depends on $ \bgee{r}$ only through rotationally invariant terms, 
we conclude that
$ [\, \Hhub, \bQ \;] = 0 $. We further know that the Hubbard
model commutes with ordinary spin whereby
$ [\, \Hhub, \bS  \; ] = 0 $.  We  can therefore conclude that
$ [\, \Hhub, \bP\; ] = 0 $, which
confirms the well known invariance under pseudospin rotations.

The operators $ T_{\pm1,0} $ are 
identical to the operators defined in Ref. \cite{macdonald}  and used
in  a perturbation expansion in $t/U$. 
It is clear from 
their definitions that $T_{\pm 1}$ couples different 
Hubbard bands defined by the sectors given by different
values of $\e{h_U}$ whereas $T_{0}$ does not.  
We observe
$ [\,T_{0,\pm 1}\,,\, \bQ\,] = 0 $.  It can be inferred that
$ [\,T_{0\pm 1}\,,\, \bP\,] = 0 $
since $ T_{0\pm 1} $ is known to commute with ordinary spin.
\newcommand{\sS}{{\cal S}}

We now define the usual canonical transformation $ \sS $ which splits the
Hamiltonian into pieces that preserve total quasicharge.
We define $ \sS = i ( \teeminus - \teeplus)  \; + \; i \; [\;\teezero,
(  \teeplus + \teeminus )\; ] $ 
and it straightforward to 
show that  $ [\, \sS, h_U\,] = ( \teeminus + \teeplus )  $, whereby 
it follows 
that
$ H_{eff} = e^{i t \sS /U } H_{hub} e^{-i t \sS / U }   =  $ 
$ H_{hub} + $ $   i \, t  [\sS,H_{hub}] / U    -  $ 
$  t^2 [\sS , [\sS, H_{hub} ]] / ( 2 U^2 ) + ...   $ 
$ \equiv    H_{q\hat{c}} + H_{q}   + O(t^3/U^2) $
splits into two terms. The first  term $ H_{q} $ is the
ordinary spin interaction in the ``spin-only'' sector
\begin{equation}
H_{q} = \jay \sum_{\e{r\rp}} (\bgee{r}\cnodot\bgee{\rp} - \quarter ) 
\end{equation}
\newcommand{\rpp}{r^{\prime\prime}}
where $ \jay = 4 t^2 / U $ and the second term $ H_{qc}$ couples
quasispin and quasicharge
\begin{multline}  \label{eq:pert}
 H_{q\hat{c}}  = \sum_{\e{r\rp}} ( U h_U + t \;   \teezero +    \frac{t^2}{4 U}
  ( n^f_r + n^f_{\rp})(\bgee{r}\bgee{\rp} - \quarter )) + \\ 
  + \frac{t^2}{4 U } \sum_{\e{r\rp\rpp}}  ( \chdop{r} \chop{\rpp} \, + \, \chdop{\rpp} \chop{r}    )\, ( 1 - 
\bgee{r}\bgee{\rp} + \bgee{r}\bgee{\rpp} +   \\ - \bgee{\rp}\bgee{\rpp}  
+i\bgee{r}\cdot(\bgee{\rp}\times\bgee{\rpp})\;.  )  
\end{multline}
The last summation runs over all three-site neighbors connected
by links. Terms of order $ t^2/U$ which are already present to order
$ t $ or $ U $ are ignored. 
According to the results of Ref. \cite{macdonald}, to all
orders in $ t/U $ the operator $ \sS $ which diagonalizes total
quasicharge is a polynomial of $ T_{0,\pm 1} $, and hence to all orders in 
perturbation theory, the transformation by $ e^{i t \sS /U } $ will preserve 
all the symmetries $ Q $ and $ P $ in this formulation.

\ssection{The spin-only sector}
Under this canonical transformation the eigenstates
of $ H_{eff} $ will map to the eigenstates of the Hubbard model represented by
the bare quasiparticle operators through the transformation $ e^{i t \sS /U  } $. 
We let $ \ket{\Psi_{eff} }$ be eigenstates of 
$ H_{eff} $ and $ \ket{\Psi}  = e^{i t \sS/U} \ket{\Psi_{eff}} $.
Adapting 
Ref. \cite{macdonald}  to our 
problem, there will be a ``zeroth Hubbard band'' or ``spin-only sector''
of solutions corresponding
to  $ n_r^f \ket{\Psi_{eff} }  $ being identically zero
so that $ H_{eff} $ reduces exactly to $H_{q}$.
We  conclude $\e{H_{q\hat{c}}}_{\Psi_{eff}}$
is
identically zero and the effective Hamiltonian of
the zeroth subband 
to order $t^2/U$ is exactly the Heisenberg model.

What can we conclude from symmetries about the properties of the zeroth
Hubbard band?  It must be remembered that 
$ N^f_{tot} = \sum_{r} n^f_r $ is not a ``good quantum
number'' i.e. is neither invariant under the symmetries
nor under $ e^{i \sS} $. We have therefore no information about
$ \e{n^f_r}_\Psi $ from symmetry arguments alone.
However, we see that $ n^f_r \ket{\Psi_{eff} } = 0 $ 
throughout the entire lattice in the entire zeroth Hubbard subband.
We can therefore conclude that
$ \e{\bP}_{\Psi_{eff}}= 0 $.
Since $ \bP $ commutes with $ \sS $, this assertion survives to
all orders in perturbation theory and  we can conclude that
$ \e{\bP}_\Psi = 0. $
In fact quasispin and ordinary spin coincide. Since
$ \e{n}=\e{1-2P_z} $ is equal to physical charge, we see that the spin-only sector
has indeed charge density corresponding to half filling.

\ssection{Extended states in the charge-only sector}
The pseudospin is given by 
$ \bP^2 = \sum_{r,\rp} \bgee{r}\cdot\bgee{\rp} \; n^f_r n^f_{\rp} $
If
there is exactly one quasicharge, we conclude from
the fact that $ \bgee{r}^2 = 3/4 $ that
$ \e{\bP^2} =  \frac{3}{4}  $.  We 
have thus  identified the lowest
nonzero quasicharge sector with the spin half representation of
pseudospin, exactly what we need to associate it with one added positive
or negative physical charge.

We can derive a set of extended charge states in the dilute limit by a mean field 
treatment of Eq. \ref{eq:pert}, replacing $\bgee{r}\cdot\bgee{\rp}=\e{\bgee{r}\cdot\bgee{\rp}}_{H_q}$.
In two dimensions, these 
have the spectrum
\begin{multline} E_k  =   \half \, U - 4 t \alpha ( \cos{ k_x} + \cos{k_y} ) 
	+  \\
   \frac{t^2}{4 U } ( ( \cos{2 k_x} + \cos{2 k_y} ) ( 1 - e_{2}) + \\
	  2 \cos{ k_x} \cos { k_y} \,  ( 1 - e_{\sqrt{2}} ))
\label{extended}
\end{multline}
where $ \alpha = |( \quarter + e_{1})|$ and where $ e_{1} =\e{\bgee{r}\bgee{\rp}}$, 
$ e_{\sqrt{2} } $ and $ e_{2} $ denote nearest neighbor 
and longer range Heisenberg correlations.
Using the numerical value $ e_{1} = -0.33972(2) $
\cite{heis} we find $ \alpha = .08972 $. 
We find a quasicharge spectrum bounded by $ U/2 \pm  8 \alpha t  + O(t^2/U) $
and shown in the shaded region of Fig. \ref{fig:fig2}.

\ssection{The Nagaoka instability and localized states}
The present calculation  makes the ``Nagaoka theorem'' natural.\cite{nagaoka}
Consider  Eq. \ref{eq:newhub} in the case where the quasispins 
are perfectly aligned along the quasispin z axis. In this case 
$ \bgee{r}\cdot\bgee{\rp} = \quarter $,
the operator $ T_{\pm 1}$ 
will be identically zero and the kinetic energy $ T_0  $ 
becomes precisely 
$  \sum_{r\rp}  \chdop{r}\chop{\rp} + \chdop{\rp}\chop{r} $.
These are quasicharges moving independently of the
ferromagnetic quasispins and are 
ordinary holes in a perfect 2-D ferromagnet which
the kinetic energy   
$  - 2 t ( \cos{ k_x a} + \cos{ k_y a } ) +  \half U $  
in the thermodynamic limit.  

The Nagaoka theorem states that as $ U \rightarrow \infty $ a single
charge causes the ground state of $ H $ to become ferromagnetic.
This Nagaoka ferromagnet with charge one has Hubbard energy 
$ E_{ferro} = \half \,  U - 4 t $, whereas the antiferromagnetic
solution with a single quasicharge has energy 
$ E_{antiferro} =    2 \Omega \jay \; (e_{1}-\quarter)  +  U/2 - 8 \alpha t
$ where $ \Omega $ is the number of lattice sites in the system.
We therefore find that  $ E_{ferro} < E_{antiferro} $ when the
inequality $   U > \,  \Omega  \,  t \,  (1+2 \alpha)/(1-2\alpha)$ holds.
We see that whether or not there is Nagaoka ferromagnetism depends
on which order we take $ U $ and $ \Omega $ to infinity.

The argument can be used to estimate
when the antiferromagnet becomes
unstable to ferromagnetic ordering. (See also Ref. \cite{phasesep} ). Let us consider
a quasicharge density of $ \rho \ll 1 $. 
We see that if  $ \rho $ obeys 
$ \rho < t/U \;(1+2 \alpha)/(1-2\alpha) $ the dilute
ferromagnet will be energetically favored over 
a collection of extended quasicharge states, essentially 
reproducing the argument by Ioffe and Larkin.\cite{phasesep}

Using the previous ideas, we shall study localized states in
a 2-D antiferromagnet.  As we have seen, a quasicharge hops freely on a ferromagnetic bond where 
$ \bgee{r}\cdot\bgee{\rp} = \quarter $.  In a mean field treatment of the
quasispins, to linear order in $t$ a quasicharge cannot hop an a bond with
antialigned quasispins  since $ \e{\bgee{r}\cdot\bgee{\rp} } = -\quarter $. 

Starting with a patch of locally Neel ordered quasispins oriented along $ \pm q^z $ 
we consider a quasispin state with nearby spins flipped relative
to the Neel state on diagonally adjacent sites of the spin-down sublattice.
We note that in the presence of a quasispin
ferromagnet oriented along $ + z $, quasicharge corresponds to physical charge
measured from half filling.

\begin{figure}[!t]
\includegraphics[width=8.5cm]{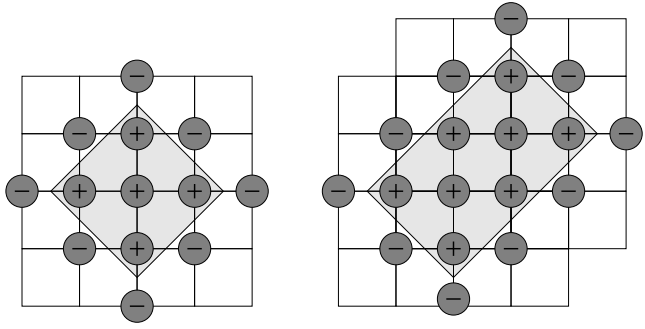}
\caption{\label{fig:fig1}(a,b)  \hspace{.2cm}
The two smallest 
quasiferromagnetic clusters is shown.
The shaded regions include the sites on which charge is approximately confined.}
\end{figure}

\begin{figure}[!t]
\vspace{0.2cm}
\includegraphics[width=8cm]{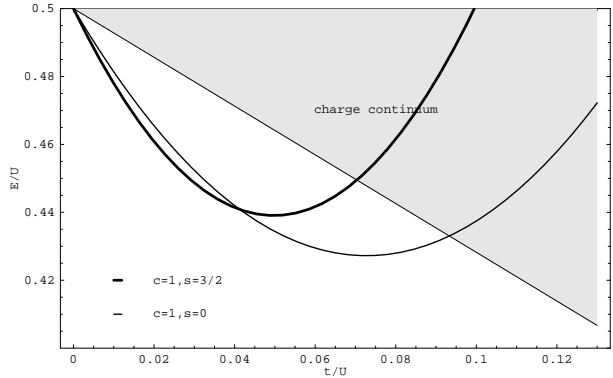}
\caption{\label{fig:fig2} Energy of the extended and 
the two simplest localized single-charge states. }
\end{figure}

The core of the region will thus
be  a
ferromagnet
surrounded by a closed boundary of
antialigned spins.  Outside this patch,
quasispins are allowed to relax,  so that far away the system will be a
Heisenberg antiferromagnet. The two smallest such configurations are
shown in Fig. \ref{fig:fig1}.

We identify the quasispin of the state by constructing the Neel state with
the same rotational symmetry around the cluster. The smallest cluster
(Fig.\ref{fig:fig1},a) has
one flipped spin thus has $Q=\half$ whereas the cluster with two flipped
spins (Fig. \ref{fig:fig1},b) has $Q=2$.  
We now add $n_f$ quasicharges to
the cluster. According to the definition of pseudospin,
in the case where quasicharge is confined to a quasiferromagnetic region
with $\bgee{r}\cdot\bgee{\rp}=\quarter+\half\delta_{r\rp}$ , it can be
shown that $\bP^2=\frac{n_f}{2}(\frac{n_f}{2}+1)$. A cluster with $n_f$
units of quasicharge will thus have pseudospin $P= \half \, n_f $ and, from
the identity $\bQ=\bS+\bP$, physical spin $S=|Q-\half n_f |$. The 
clusters with no quasicharge have pseudospin zero and physical spin
equal to quasispin, whereas the charge one cluster will have pseudospin
$ P=\half $ and spin $ S = Q-\half $.

The energies of these states are approximate but the 
spin and pseudospin quantum numbers are exact.
As quasicharge is added in the cluster,
quasispin, spin and pseudospin continue to be good quantum numbers.
Since  quasispin is conserved upon adding quasicharge
each such configuration has a well defined value of quasispin
independent of physical spin or charge and can be calculated
in the effective theory. Quasicharge, however, is not a good 
quantum number, and will change upon transforming back the
the physical state with $ e^{i t {\cal S } /U  } $.

We estimate the energy by keeping only
the  part of the kinetic energy  linear in $ t $ 
and ignore the spin exchange energy beyond the antiferromagnetic boundary of the cluster. 
For the smallest $Q=\half$ cluster we see sixteen bonds 
that are strongly affected by the localized charge.
Four bonds are ferromagnetic with zero energy and twelve bonds are antiferromagnetic with 
energy $ \jay \times ( - \half  ) $, i.e.  they retain
about 85\% of the energy of the unaffected Neel state $ \jay \times (e_1 - \quarter)\approx -.59 \jay $. 

\newcommand{\Es}[3]{ {\phantom 1}^{#1}E^{#2}_{#3}  }
\newcommand{\thalf}{\scriptscriptstyle \frac{1}{2}}
\newcommand{\threehalf}{\frac{3}{2}}

The energy of these states is compactly written with
the spectroscopic-like notation 
$ \Es{Q}{c}{S} $ where $ Q $ is the quasispin,
$ c=2P^z $ is the charge relative to half filling and $S$ is the spin. 
Thus $ \Es{\thalf}{0}{\thalf} $ is the energy of the the smallest 
uncharged cluster measured with respect to the Heisenberg ground state. We find
$ \Es{\thalf}{0}{\thalf} = \jay ( 12 \times (-\half) + 16 \times |e_1+\quarter| )  \approx  13.74\; t^2/U $. 
The five 
kinetic energy eigenvalues of this cluster are  $ t ( -2,0,0,0,2) $. 
We thus find $ \Es{\thalf}{1}{0} =U/2+\Es{\thalf}{0}{\thalf} -2t \approx  U/2+\;13.74\; t^2/U -2t$.
For the $Q=2$ cluster we find  a kinetic energy for a single charge
$ -\sqrt{6} t $; there are 16 antiferromagnetic bonds
and 8 ferromagnetic bonds giving 
$ \Es{2}{1}{\threehalf} = U/2+\Es{2}{0}{2}  - \sqrt{6} t \approx U/2+\; 24.61\;t^2/U-2.45 t$.
The energy of these states are plotted in Fig. \ref{fig:fig2} and compared to the energy of the band of extended states 
from Eq. \ref{extended}. 
We find that for large $U$ the clusters with localized charge are energetically favored over the extended states. 
As $ U $ becomes larger, larger clusters lower the energy in accordance with
the Nagaoka theorem.
\ssection{Conclusions}
We have demonstrated an exact local mapping from a Fermionic many-body
system to system of interacting Bosons and Fermions  that does not
introduce constraints.
The mapping provides a paradigm for a non-Fermi liquid behavior for
correlated Fermions valid in any dimension.
Applied to the 2-D Hubbard model the transformation makes precise earlier attempts 
that relied on approximations and constraints to discuss spin and charge 
as composite operators. 
Our calculations 
supports the idea of low energy localized states where spin 
is bound to charge in the limit of large $U$.


\begin{thebibliography}{8}
\bibitem{LutherEmery}
V.J. Emery,{\it Theory of the one-dimensional electron gas}, in ``Highly Coducting One-Dimensional
Solids'', edited by J.T. Devreese, R.P.Evrard and V.E. van Doren, 327 (Plenum, New York, 1979);
J. Solyom, {\it Adv. Phys.  }{\bf 28 }, 201-303 (1979).
\bibitem{fradkin}
P. Coleman, {\it Phys. Rev. B. }{\bf 29,} 3035, (1983);
See, E. Fradkin, {\it Field Theories of Condensed matter physics}, 
( Addison-Wesley, 1991 ) and references therein.
\bibitem{mele}
Similar strictly canonical transformations 
that do not map between fermionic and bosonic operators
have been discussed by S. \"Ostlund, E.J. Mele, Phys. Rev. B. {\bf 44 }, 12413 (1991);
S. \"Ostlund, T.H. Hansson, A. Karlhede, Phys. Rev. B{\bf 71}, 165121 (2005)
\bibitem{gunn}
After this work was completed, we have found out that that 
the transformation appears in unpublished work by J.I. 
Chandler and J.M.F. Gunn, cond-mat/9702009.
\bibitem{shiba72}
H. Shiba, {\it Prog. Theor. Phys} {\bf 48 }, 2171 (1972); 
C. N. Yang and S. C. Zhang, Mod. Phys. Lett. B 4, 759 (1990);
Shoucheng  Zhang, Phys. Rev. Lett. 65, 120 (1990);
V. J. Emery, Phys. Rev. B 14, 2989 (1976); 
A.B. Eriksson, T. Einarsson and S. \"Ostlund,  Phys. Rev.  {\bf B }52 , 3662 (1995) 
\bibitem{nagaoka}
Y. Nagaoka, {\it Solid State Comm.}{\bf 3}, 409 (1965); Phys. Rev.
{\bf 147}, 392 (1966); D.J. Thouless, {\it Proc. Roy. Soc. }{\bf 86},
893 (1965).
\bibitem{macdonald}
A.H. MacDonald, S.M. Girvin and D. Yoshioka, {\it Phys. Rev. B }{\bf 37},
9753 (1988);
J.-Y.P. Delannoy, M.J.P. Gingras, P.C.W. Holdsworth and A.-M.S. Tremblay,
cond-mat/0412033.
\bibitem{heis}
E. Manousakis, Rev. Mod. Phys. {\bf 63 } 1, (1991).
\bibitem{phasesep}
P.B. Visscher, Phys. Rev. B {\bf 10}, 943 (1974);
L.B. Ioffe and A. I. Larkin, Phys. Rev. B. {\bf 37}, 5730 (1988).
\end{thebibliography}
\end{document}